\documentclass{PoS}

\usepackage{amsmath}
\usepackage[font=small]{caption}
\usepackage{subcaption}
\usepackage{url}

\title{Atmospheric monitoring and array calibration in CTA using the Cherenkov Transparency Coefficient}

\ShortTitle{Atmospheric monitoring and array calibration in CTA using the CTC}

\author{\speaker{Stanislav Stefanik}$^a$, 
Raquel de los Reyes$^{b}$\footnote{Now with the German Aerospace Center (DLR), Earth Observation Center (EOC), D-82234, Wessling, Germany}, 
Dalibor Nosek$^{a}$ for the CTA Consortium\thanks{http://cta-observatory.org}\\
\llap{$^a$}Charles University, Faculty of Mathematics and Physics\\
V Holesovickach 2, 180 00 Prague, Czech Republic\\
\llap{$^b$}Max-Planck-Institut f\"ur Kernphysik, Saupfercheckweg 1, 69117 Heidelberg, Germany \\
E-mail: \email{stefanik@ipnp.troja.mff.cuni.cz}}

\abstract{
The Cherenkov Telescope Array (CTA) will be the next generation observatory employing different types of Cherenkov telescopes for the detection of particle showers initiated by very-high-energy gamma rays.
A good knowledge of the Earth's atmosphere, which acts as a calorimeter in the detection technique, will be crucial for calibration in CTA.
Variations of the atmosphere's transparency to Cherenkov light and not correctly performed calibration of individual telescopes in the array result in large systematic uncertainties on the energy scale.
The Cherenkov Transparency Coefficient (CTC), developed within the H.E.S.S. experiment, quantifies the mean atmosphere transparency ascertained from data taken by Cherenkov telescopes during scientific observations.
Provided that atmospheric conditions over the array are uniform, transparency values obtained per telescope can be also used for the calibration of individual telescope responses.
The application of the CTC in CTA presents a challenge due to the greater complexity of the observatory and the variety of telescope cameras compared with currently operating experiments, such as H.E.S.S.
We present here the first results of a feasibility study for extension of the CTC concept in CTA for purposes of the inter-calibration of the telescopes in the array and monitoring of the atmosphere.
}

\FullConference{35th International Cosmic Ray Conference\\
         10-20 July, 2017\\
         Bexco, Busan, Korea} 

\begin{document}

%***********************************************************************************
% INTRODUCTION
%***********************************************************************************

\setcounter{page}{2}

\section{Introduction}
\vspace{-0.2cm}

The Cherenkov Telescope Array (CTA) is a ground-based very high energy (VHE) gamma-ray observatory in the pre-construction phase~\cite{OngICRC2017}. 
It will consist of two arrays on both Earth's hemispheres including not only different optical systems and detector hardware, but also different sizes of telescopes. 
A complete strategy on the calibration of the full array~\cite{CettinaICRC2017} as well as the atmosphere~\cite{EbrICRC2017} is currently under development in CTA.
The Cherenkov Transparency Coefficient (CTC) is included within the calibration strategy of both:
\vspace{-0.2cm}
\begin{itemize}
    \setlength\itemsep{-0.2em}
    \item Removing the system dependency of the stereo trigger rates, the CTC will depend only on the atmospheric extinction of the Cherenkov light emitted by the air showers.
    \item Utilizing the optical throughput dependency of the CTC, normally neutralized by the throughput estimated with the muons, will allow the CTC to monitor the variations of the detection efficiency of the telescopes, alternatively to their assessment through the muon analysis.
\end{itemize}
\vspace{-0.2cm}

In the next sections, we will describe the steps to apply the results of the  H.E.S.S.~collaboration~\cite{Hahn} and extend them for their use in more complex systems, like the CTA. 
This study makes use of Monte Carlo (MC) simulations of protons observed by a candidate array of telescopes located at the northern CTA site (CTA-N) at La Palma~\cite{MC}.
The array consists of 4 large-sized (LST) and 15 medium-sized (MST) telescopes with the mirror dish diameters of 23~m and 12~m, respectively~\cite{HassanICRC2017}.

In Sec.~\ref{Sec:Stereo}, we will give a brief introduction about the concepts involved in the CTC, including the discussion of the effects of the array layout geometry (Sec.~\ref{Sec:Geometry}), the influence of the Earth's magnetic field (Sec.~\ref{Sec:Bfield}) and the array hardware (Sec.~\ref{Sec:Hardware}).
In Sec.~\ref{Sec:Intercalibration}, we will describe the role of the CTC in the inter-calibration of the CTA telescopes of the same size.
Sec.~\ref{Sec:Atmosphere} deals with the use of the CTC for the monitoring of the atmospheric transparency.
In Sec.~\ref{Sec:Conclusions}, a summary of the status of this feasibility study will be given together with future steps to be undertaken.

%***********************************************************************************
% STEREO TRIGGER RATES
%***********************************************************************************

\vspace{-0.2cm}
\section{Trigger rates of Cherenkov telescopes}
\label{Sec:Stereo}
\vspace{-0.2cm}

During the scientific observations of the CTA, the telescopes will record events seen by at least two telescopes (stereo trigger events).
In contrast with the single telescope triggers, the stereo requirement will partially eliminate random fluctuations due to the night sky background and accidental triggers.
In a first approximation, the rate of triggered events is mainly determined by the lowest detectable energy of cosmic rays (energy threshold).
This energy threshold depends inversely on the detection efficiency of the stereo partners (their effective area).
Variations of the effective area depend on the transparency of the atmosphere $T = e^{-\mathrm{AOD}}$, where AOD is the aerosol optical depth.

Based on these assumptions, the H.E.S.S.~experiment defined the transparency coefficient~\cite{Hahn} as $CTC = (N\cdot k_{N})^{-1}\cdot\sum_{i}{ R_{i}^{\frac{1}{1.7}}\cdot( \mu_{i} \cdot g_{i} } )^{-1}$, where the sum runs over each of the $N$ active telescopes, $R_{i}$ is the stereo trigger rate at zenith of all events triggering the $i$-th telescope together with at least one other telescope, $\mu_{i}$ is the muon-estimated normalised optical throughput, $g_{i}$ is the average pixel gain and $k_{N}$ accounts for layout-related changes.

The normalization factor $k_{N}$ also includes a dependency of the system rate on the distance between telescopes ($D$), their orientation with respect to the shower ($\beta$) and the effects of the Earth's magnetic field ($\vec{B}$).
For arrays with tens of telescopes, like CTA, this results in a vast number of possible realizations which cannot be straightforwardly included in the H.E.S.S.~formula.
In order to account for all possible dependencies in $k_{N}$, the stereo trigger rate $R$ used in the definition of the CTC must be modified (by a function $F$) to ensure the independence of the transparency estimate $\tau$ for CTA from hardware and observation-related quantities:
\begin{equation}
    \label{Eq:transparency}
        \tau (\mathrm{AOD}) = R \left( \mathrm{AOD}, D, \theta, \beta, \vec{B}, \varepsilon \right) \cdot F^{-1}\left( D, \theta, \beta, \vec{B}, \varepsilon  \right).
\end{equation}
where $\theta$ is the zenith angle (see Fig.~\ref{Fig:coordinates}) and $\varepsilon$ describes the hardware dependency studied in Sec.~\ref{Sec:Hardware}.

%***********************************************************************************
% GEOMETRICAL CONFIGURATION
%***********************************************************************************

\subsection{Geometrical configuration}
\label{Sec:Geometry}

\setlength\belowcaptionskip{-1ex}

\begin{figure}[!t]
\centering
\begin{minipage}{.48\textwidth}
  \centering
  \includegraphics[width=0.78\columnwidth]{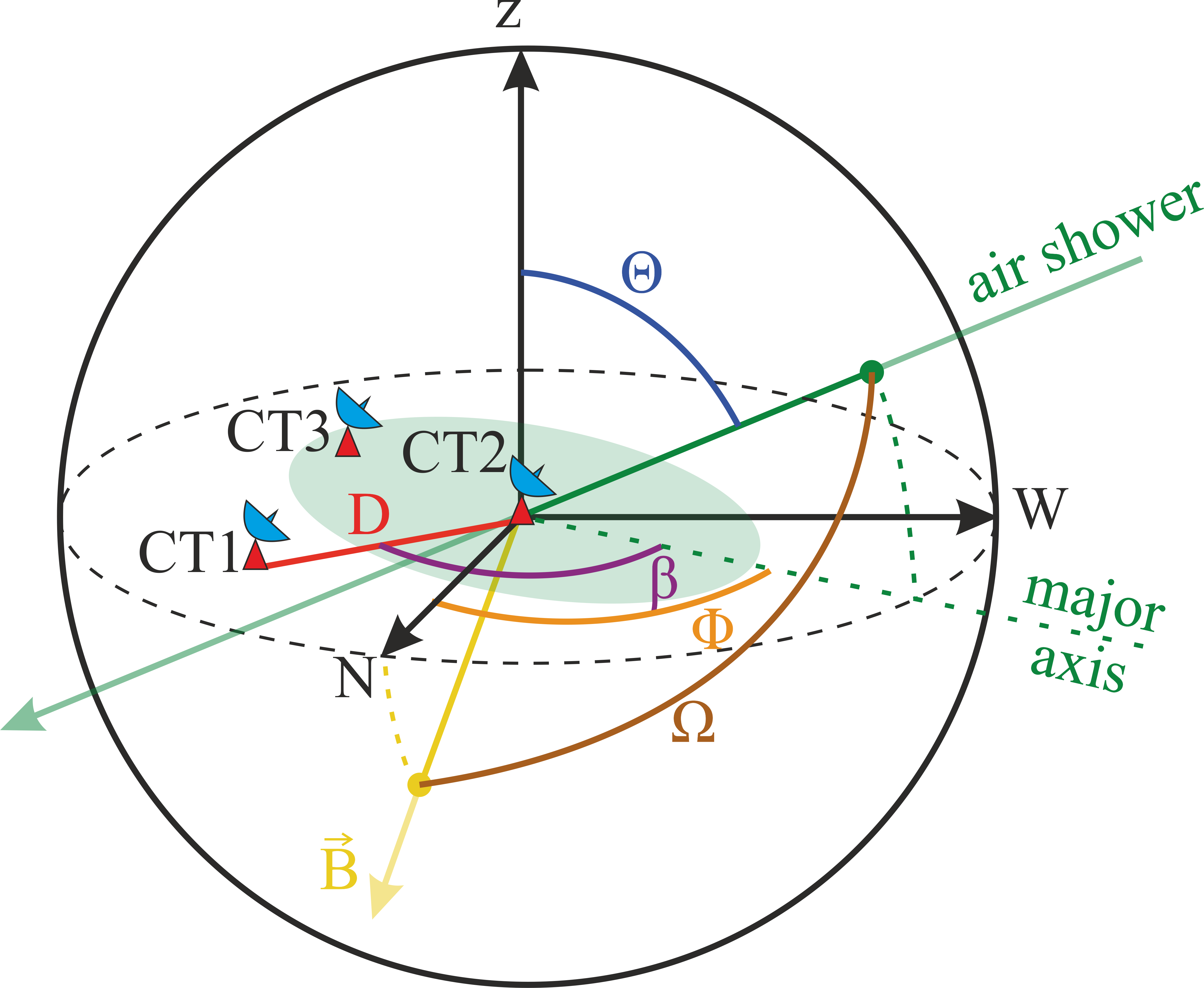}
        \captionof{figure}{
			Reference coordinate system. $\beta$ is the angle between the line joining the telescopes (CT1 and CT2 at a distance $D$) and the shower azimuth direction ($\phi$).
			$\Omega$ is the angle between the shower and the magnetic field $\vec{B}$  direction.
                        }
    \label{Fig:coordinates}
\end{minipage}
\hfill
\begin{minipage}{.48\textwidth}
  \centering
  \includegraphics[width=\columnwidth]{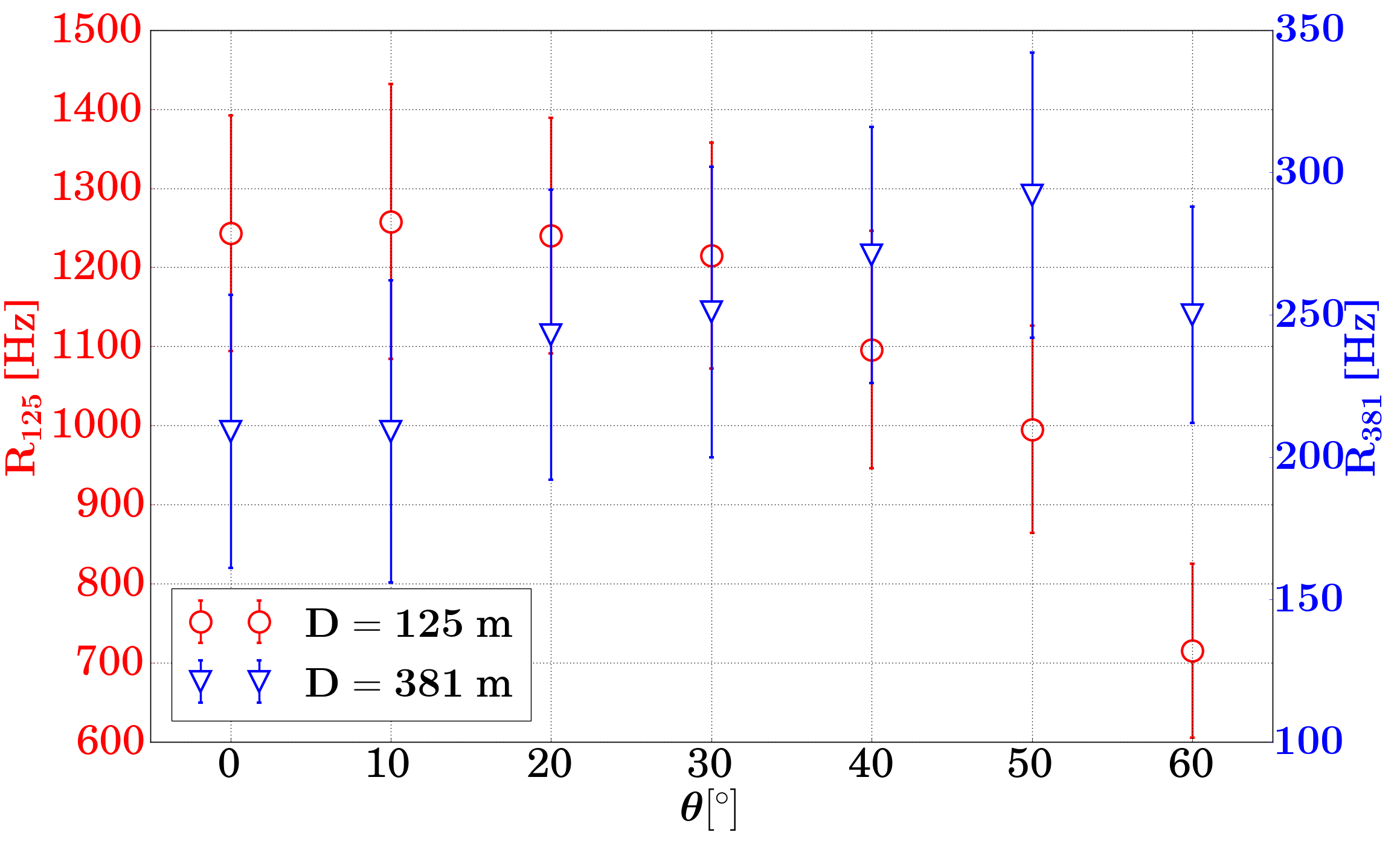}
  \captionof{figure}{
			Stereo trigger rate of two MSTs vs shower direction for fixed $\phi=180^{\circ}$.
			Colours and vertical axes represent values of trigger rate registered by the telescopes separated by distances 125~m (red) or 381~m (blue).
			}
  \label{Fig:zenith}
\end{minipage}
\end{figure}

In this section, we examine the relationship between the trigger rates, the distance $D$ between telescopes and their relative position $\beta$ with respect to the shower direction given by $\theta$ and $\phi$ (see Fig.~\ref{Fig:coordinates}).
The study utilizes 2-telescope trigger rates obtained from a set of MC simulations~\cite{MC} of proton showers for the CTA-N site.
The shower directions were given by $\theta \in [0^{\circ},60^{\circ}]$ with $\phi = 0^{\circ}$ or $180^{\circ}$.
The positions of telescopes were chosen such that the different pairs were aligned with respect to the showers at angles $\beta =0^{\circ}$, $30^{\circ}$, $60^{\circ}$ or $90^{\circ}$.
In order to cover a larger range of telescope distances, MSTs were simulated at 9 different positions for each orientation $\beta$.

\begin{figure}[!t]
	\begin{center}
		\includegraphics[width=0.49\columnwidth]{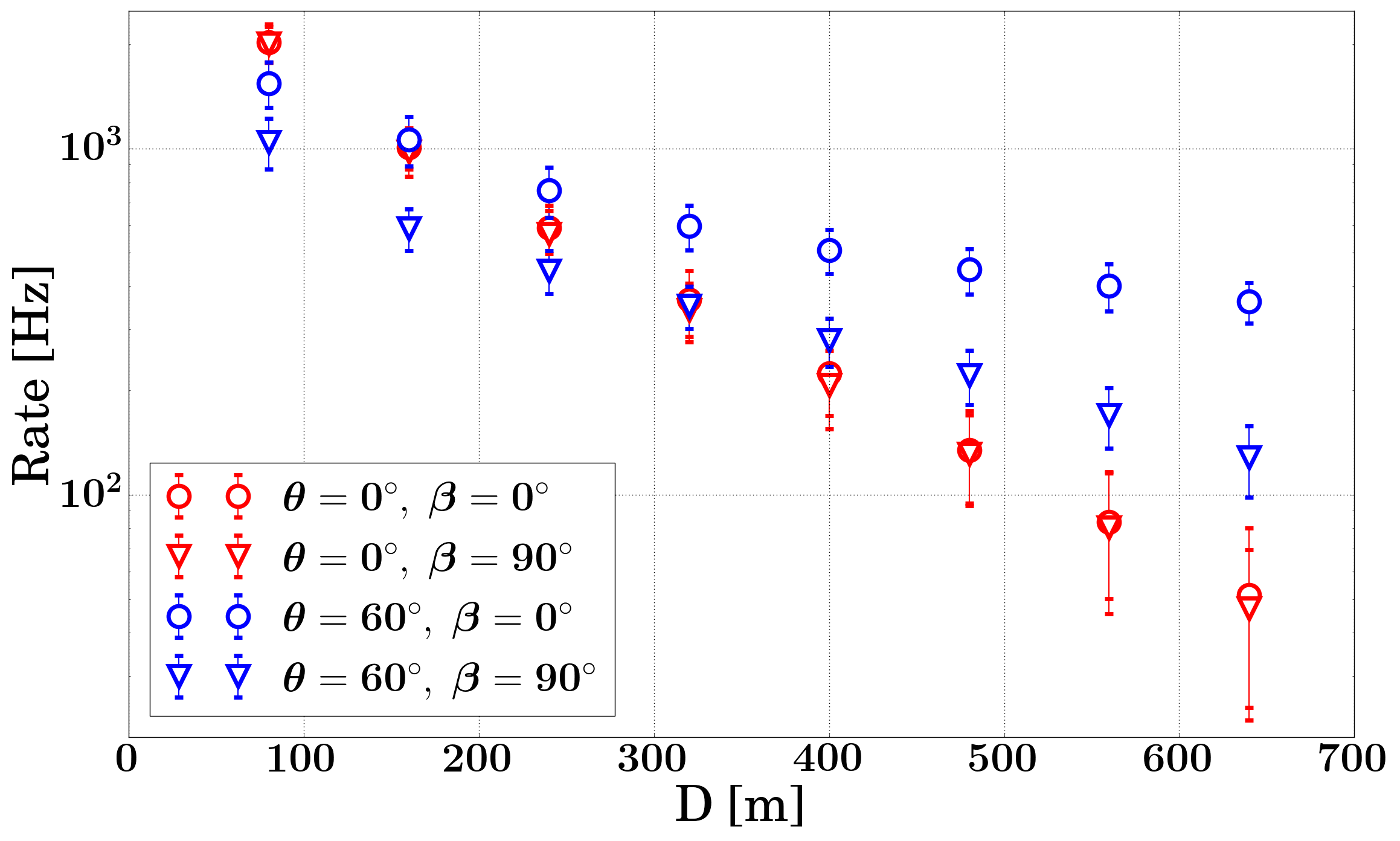}
		\hfill
		\includegraphics[width=0.49\columnwidth]{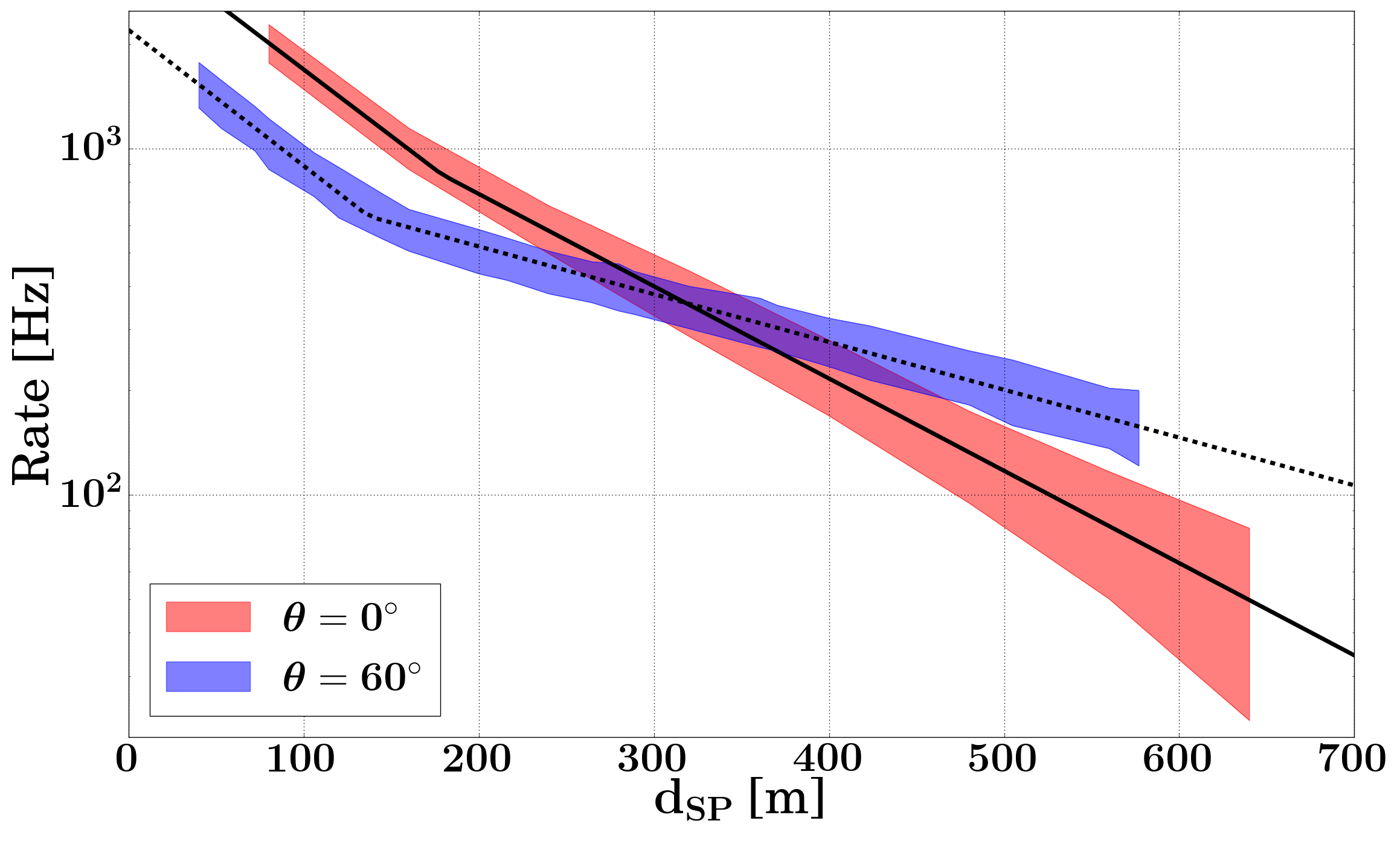}
		\caption{
			Left: trigger rate of two MSTs vs the telescope separation~($D$) and their relative orientation ($\beta$, different markers) with respect to the shower direction given by the zenith angle ($\theta$,~different colours).
			Right: solid and dashed lines show the fit results of Eq.(\ref{Eq:rate_fit}) to the data vs $d_{\mathrm{SP}}$ for $\theta = 0^{\circ}$ (red) and $60^{\circ}$ (blue), respectively.
			$1\sigma$ statistical uncertainties (contours) of all rates from the left plot with the same $\theta$ are shown vs $d_{\mathrm{SP}}$.
		}
		\label{Fig:ZA-AZ_noMF}
	\end{center}
\end{figure}

Since the telescope trigger efficiency decreases with the distance from the shower impact point on the ground~\cite{KonradB}, the stereo rate is reduced for larger separations between the detectors ($D$).
This is shown in Fig.~\ref{Fig:zenith}, where for $\theta < 10^{\circ}$, the rate for telescopes at 381~m distance (blue) is $\sim 17\%$ of the rate at 125~m (red).
The effective area increases with $\theta$ roughly as $\propto 1 / \mathrm{cos{\theta}}$ and the more advantageous configurations are those with larger separations of telescopes.
These effects are coupled and their combined impact on the stereo rate is a matter of the separation $D$ and orientation~$\beta$ of telescopes with respect to the shower direction ($\theta$, $\phi$), illustrated in the left plot in~Fig.~\ref{Fig:ZA-AZ_noMF}.

\setlength\belowcaptionskip{-4ex}

For showers incident from the zenith ($\theta=0^{\circ}$, red markers in the left plot in Fig.~\ref{Fig:ZA-AZ_noMF}), the Cherenkov pool on ground is roughly circular and there is no dependence of rates on the telescope alignment relative to the shower direction ($\beta$).
At~higher zenith angles ($\theta=60^{\circ}$), the pool is an ellipse and telescope pairs along its major axis (CT2, CT3 in~Fig.~\ref{Fig:coordinates}; red circles in Fig.~\ref{Fig:ZA-AZ_noMF}, left) trigger in coincidence more likely than the pairs which are oriented in an orthogonal direction (CT1, CT2; blue triangles in Fig.~\ref{Fig:ZA-AZ_noMF}, left), although their separations may be the same.

The dependence on $\beta$ and $D$ can be eliminated if the rates are examined in terms of telescope separations projected in the shower plane $d_{\mathrm{SP}}$, illustrated in the right plot in~Fig.~\ref{Fig:ZA-AZ_noMF} for all data with different $\beta$ ($0^{\circ}$ and $90^{\circ}$) combined into sets according to $\theta$.
Unlike the fixed separation of telescopes $D$ in the ground plane, the distance in the shower plane is a function of the pointing of the telescopes: $d_{\mathrm{SP}} (\theta,\beta) =  D . \sqrt{ 1 - \sin^{2}{\theta} . \cos^{2}{\beta} }$.
It follows that $d_{\mathrm{SP}}=D$ when $\theta = 0^{\circ}$ (compare the red markers and contours in Fig.~\ref{Fig:ZA-AZ_noMF}) or $\beta = 90^{\circ}$ (compare the triangles and contours).

The geometrical dependence of the rate $R_{\mathrm{Fit}}(d_{\mathrm{SP}})$ can be fit by the functions (lines in Fig.~\ref{Fig:ZA-AZ_noMF})
\begin{equation}
    \label{Eq:rate_fit}
        R_{\mathrm{Fit}} (d_{\mathrm{SP}}) =
        \begin{cases}
            A_0 \cdot e^{A_1 \cdot (d_{\mathrm{SP}} - A_3)}, & \text{if}\ d_{\mathrm{SP}}<A_3 \\
             A_0 \cdot e^{A_2 \cdot (d_{\mathrm{SP}} - A_3)}, & \text{if}\ d_{\mathrm{SP}}>A_3.
         \end{cases}
\end{equation}
While this effective description removes the dependence on the distance and relative alignment of telescopes, the coefficients $A_{i}$ still depend on the zenith angle.
Functions $A_{i} (\cos{\theta})$ were found by fitting the values of $A_{i}$ obtained from fits of the trigger rate to the Eq.(\ref{Eq:rate_fit}) for six values of $\theta$ in the range $[0^{\circ},60^{\circ}]$.
The fit results provide four sets of look-up parameters (Table~\ref{Tab:fit}) which together with Eq.(\ref{Eq:rate_fit}) allow to estimate the correction $F\left( D, \theta, \beta \right)$ of the stereo trigger rate for the geometrical configuration, leaving Eq.(\ref{Eq:transparency}) as $\tau = \tau(\mathrm{AOD}, \vec{B}, \epsilon)$.
Note that a similar study was performed with the LSTs providing equivalent results.

%***********************************************************************************
% GEOMAGNETIC FIELD
%***********************************************************************************

\subsection{Earth's magnetic field}
\label{Sec:Bfield}

The Earth's magnetic field also affects the distribution of Cherenkov light on the ground~\cite{KonradB}.
Here, we investigate these effects using the MC set of Sec.~\ref{Sec:Geometry}.
In addition to this we include another MC set in this study, in which the magnetic field intensity was changed from $|\vec{B}| \equiv B\simeq 0~\mu$T to $38.7~\mu$T, consistently with the CTA-N site\footnote{\url{https://ngdc.noaa.gov/geomag-web/#igrfwmm}}.
The effects of the magnetic field are compared for the configurations with $(\theta = 20^{\circ}, \phi = 180^{\circ})$ and $(\theta = 50^{\circ}, \phi = 0^{\circ})$, corresponding to an angle between the vector $\vec{B}$ and the shower direction $\Omega$ (see Fig.\ref{Fig:coordinates}) of $5^{\circ}$ and $72^{\circ}$, respectively.

\begin{figure}[!t]
	\begin{center}
		\includegraphics[width=0.49\columnwidth]{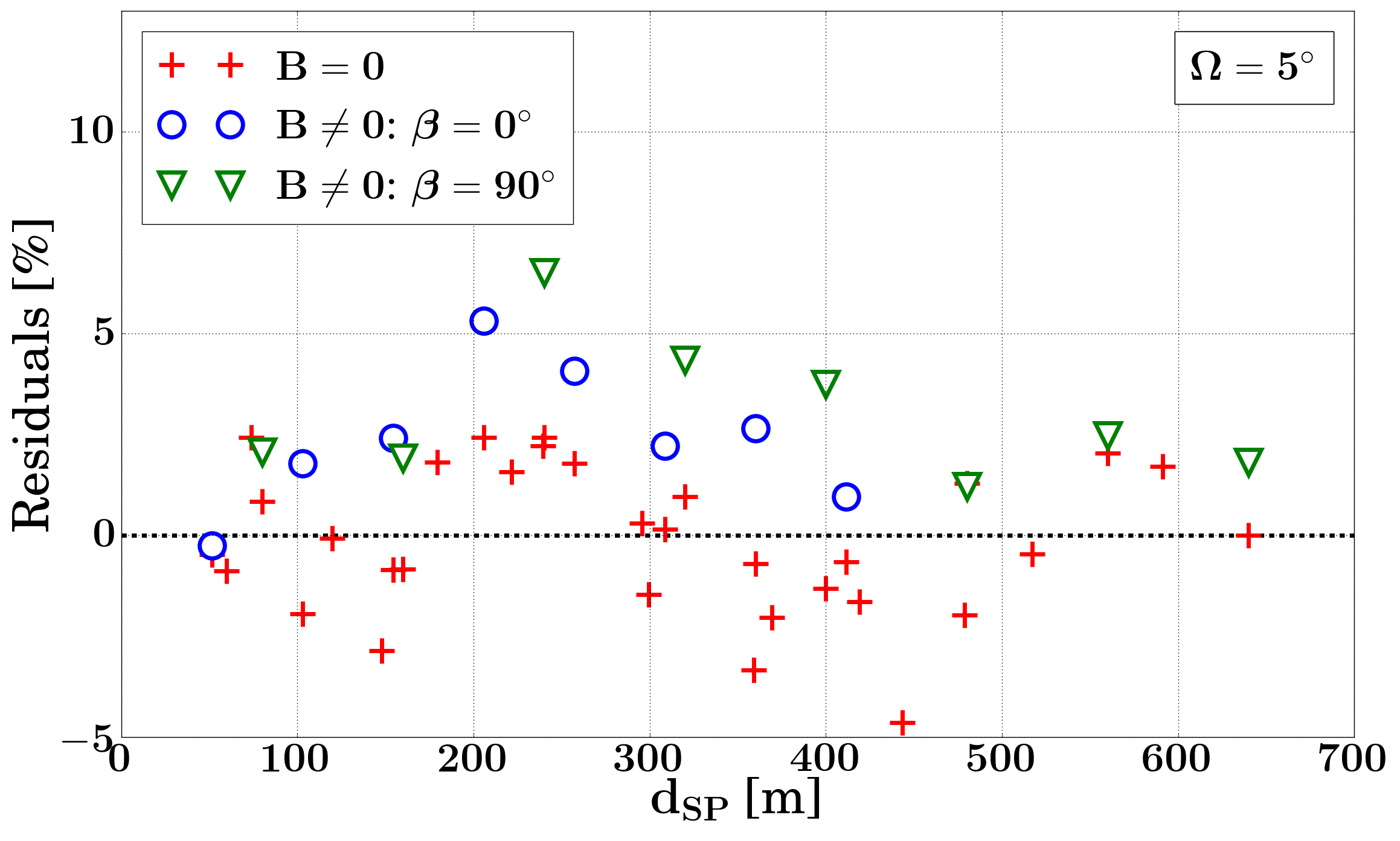}
		\hfill
		\includegraphics[width=0.49\columnwidth]{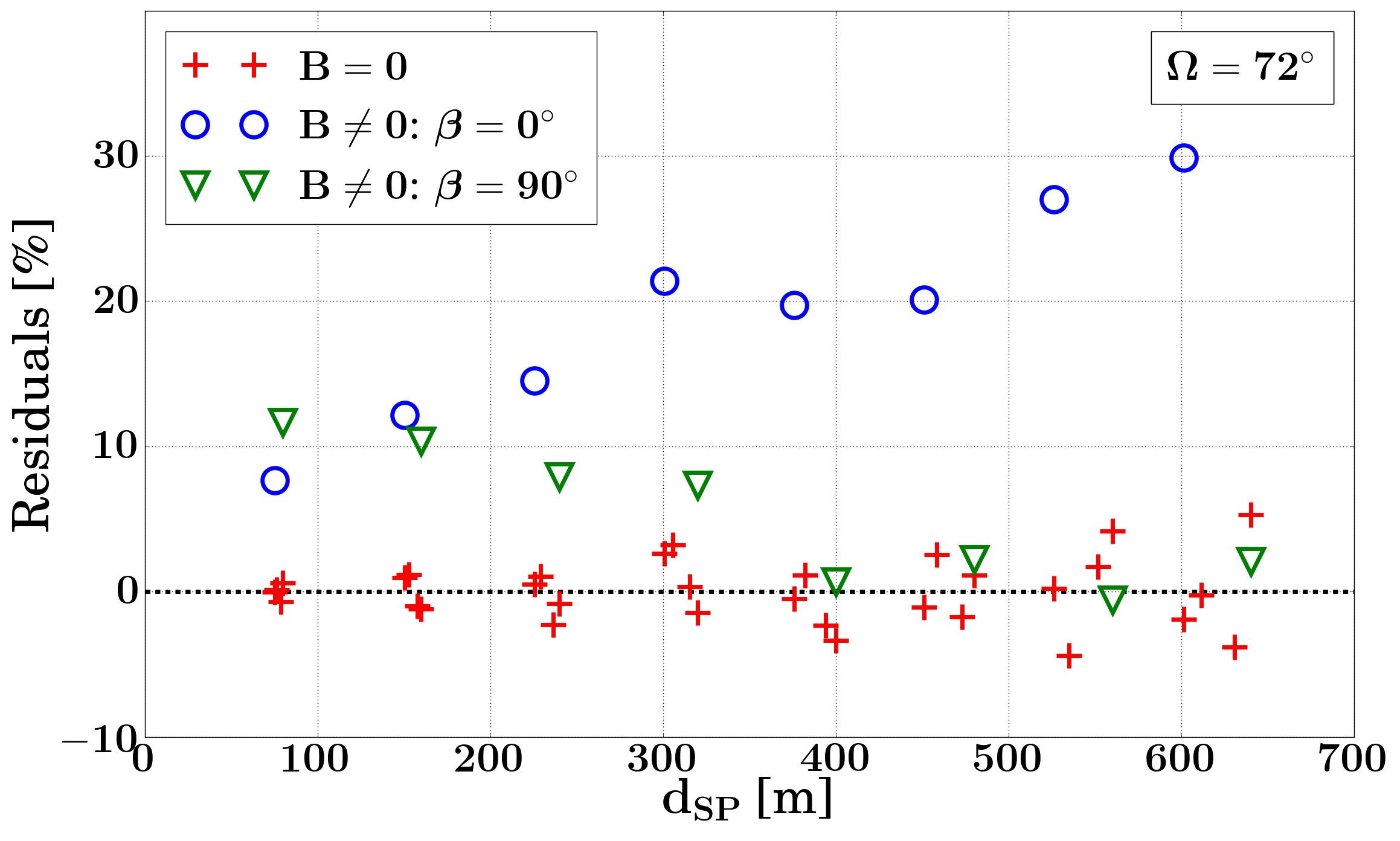}
		\caption{
			Residuals $(R_{\mathrm{Fit}}-R_{\mathrm{True}}) / R_{\mathrm{True}}$ between the simulated trigger rate ($R_{\mathrm{True}}$) and the fit value ($R_{\mathrm{Fit}}$) from Eq.~(\ref{Eq:rate_fit}) which includes only the geometrical effects ($B\simeq0$).
			Red crosses are the residuals for data with $B\simeq0$.
			Circles and triangles correspond to data with $B\neq0$ for different telescope alignment $\beta$.
			Angle~$\Omega$ (Fig.~\ref{Fig:coordinates}) is $5^{\circ}$ and $72^{\circ}$ for the left and right plot, respectively.
		}
		\label{Fig:residuals}
	\end{center}
\end{figure}

Using Eq.~(\ref{Eq:rate_fit}) and Table~\ref{Tab:fit}, we estimated the trigger rate $R_{\mathrm{Fit}}$ for data with $B\simeq 0$, accounting only for the geometrical effect.
The residuals between $R_{\mathrm{Fit}}$ and the simulated rate $R_{\mathrm{True}}$ were calculated for the data with $B\simeq 0$ and $B\neq 0$, illustrated in~Fig.~\ref{Fig:residuals} as a function of the distance $d_{\mathrm{SP}}$.
For the sake of simplicity, the statistical error bars are omitted in Fig.~\ref{Fig:residuals}.
These uncertainties rise with the distance from roughly $10\%$ to as much as $20\%$ and $40\%$ for $\Omega=5^{\circ}$ and $\Omega=72^{\circ}$, respectively.

For the shower directions nearly parallel with the magnetic field ($\Omega = 5^{\circ}$, left plot in Fig.~\ref{Fig:residuals}), the Lorentz force exerted on the particles is negligible and the trigger rates in the magnetic field (circles and triangles) are expected to be consistent with the case $B \simeq 0$ (crosses).
The deviations of $R_{\mathrm{True}}$ from the geometrical fit $R_{\mathrm{Fit}}$ are randomly distributed and are independent of $d_{\mathrm{SP}}$ and~$\beta$.
The RMS of the residuals for $B \simeq 0$ is at the level of $2\%$ for any value of $\Omega$.

For a non-zero magnetic field in a non-parallel orientation relative to the shower direction ($\Omega = 72^{\circ}$, right plot), the mean deviation of $R_{\mathrm{Fit}}$ from $R_{\mathrm{True}}$ differs by $\sim 10\%$ (circles and triangles) compared to the case $B\simeq0$.
The residuals also show a tendency with $d_{\mathrm{SP}}$ depending on the relative alignment $\beta$ of the telescopes with respect to the shower direction.
In the studied configuration, the deflection of particles in the magnetic field causes the Cherenkov pool to broaden roughly orthogonally to the shower and magnetic field direction.
A telescope alignment parallel to the shower direction (blue circles) is then less favourable than the perpendicular orientation (green triangles), as opposed to the instances with a small magnetic field effect (red crosses). 

A correction of the trigger rates for the magnetic field effects requires to take into account both the magnitude and the direction of the Lorentz force for all possible pointings of telescopes and their distances ($d_{\mathrm{SP}}$).
However, the residual distributions are rather flat for examined angles $\Omega$ up to $d_{\mathrm{SP}} \approx 200$~m (Fig.~\ref{Fig:residuals}).
The $\vec{B}$--dependence in Eq.(\ref{Eq:transparency}) may be neglected by imposing a cut on the maximum separation of telescope pairs which will be considered in the calibration.
The systematic uncertainties due to the magnetic field are then expected to be within the residual RMS for $B\simeq0$.
For a better correction we need more MC simulations at different $\Omega$.

%***********************************************************************************
% HARDWARE DEPENDENCE
%***********************************************************************************

\subsection{Hardware dependence}
\label{Sec:Hardware}

\setlength\belowcaptionskip{-1ex}

\begin{figure}[!t]
\centering
\begin{minipage}{.48\textwidth}
	\begin{center}
		\includegraphics[width=\columnwidth]{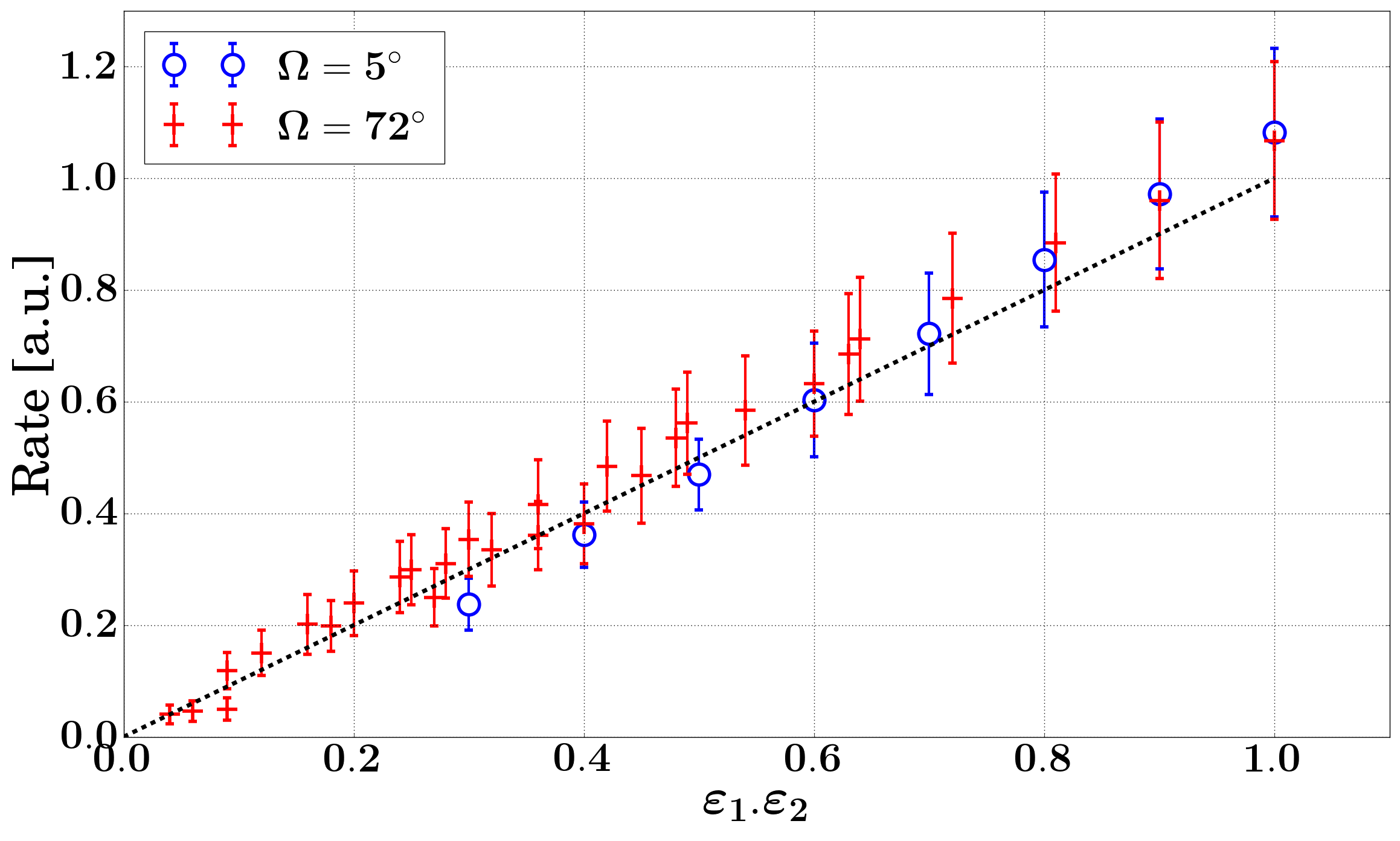}
		\caption{
		    Trigger rate of two MSTs with $d_{\mathrm{SP}} = 150$~m vs the mirror degraded efficiencies $\varepsilon_1$ and $\varepsilon_2$ for $\Omega = 5^{\circ}$ and $72^{\circ}$ (Fig.~\ref{Fig:coordinates}).
			Rates were corrected for the geometric and magnetic field effects and normalised by the value of $R_{\mathrm{Fit}}$ in Eq.~\ref{Eq:rate_fit} for $\varepsilon_{1}=\varepsilon_{2}=1$.
			The line represents $R = \varepsilon_{1} \cdot \varepsilon_{2}$.
		}
		\label{Fig:Rate_Eff}
	\end{center}

	\begin{center}
		{\small
	  \begin{tabular}{l c c c}
		\hline
		$P$ & $p_0$ & $p_1$ & $p_2$ \\
		\hline
		$A_0$ & $934\pm14$ & $15.9\pm9.9$ & $1.13\pm0.09$  \\
		$A_2$ & $(7.4\pm0.4)\text{e}^{-3}$ & $-3.7\pm0.4$ & $0.58\pm0.03$  \\
		$A_3$ & $184\pm9$ & $-6.5\pm2.2$ & $0.48\pm0.04$  \\
		\hline
	\end{tabular}	  
	}
	\captionof{table}{
	  	Fit results of the parameters of the Eq.(\ref{Eq:rate_fit}) to the relation $P = p_0 \cdot \left[ 1 + \exp(p_1 \cdot (\cos{\theta}-p_2)) \right]^{-1}$.
	  	$A_1$ can be described by the expression $p_0 \cdot [ \cos{(p_1\cdot \theta - p_2)} ] + p_3$, where $p_0 = (5.9\pm0.5)\text{e}^{-4}$, $p_1 = 6.2\pm0.3$, $p_2 = 53.9\pm7.4$, $p_3 = (8.63\pm0.03)\text{e}^{-3}$ and $\theta$ is given in degrees.
	  }
    \label{Tab:fit}
    \end{center}
	
	\end{minipage}
	\hfill
\begin{minipage}{.48\textwidth}
  \centering
  \includegraphics[width=\columnwidth]{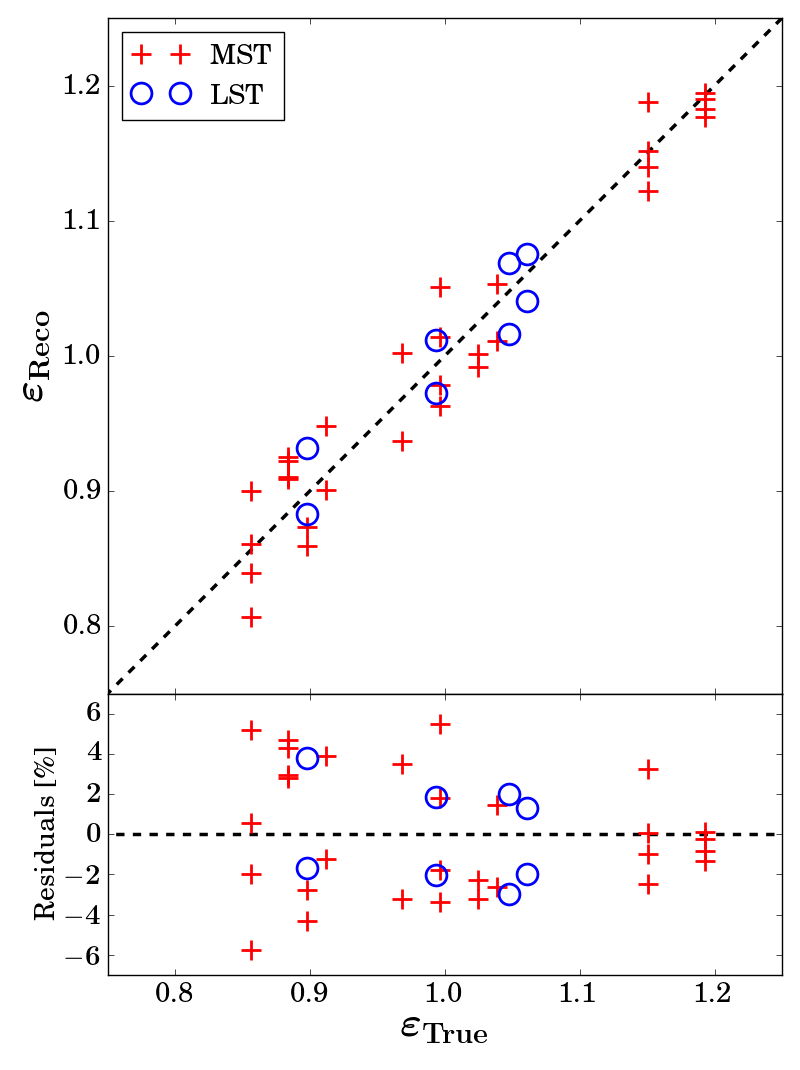}
  \captionof{figure}{
                    Optical efficiencies recovered from minimisation ($\varepsilon_{\mathrm{Reco}}$) compared to the values used as input for MC simulations ($\varepsilon_{\mathrm{True}}$).
                    Red and blue markers represent the results for MST and LST sub-system, respectively.
                    The dashed line corresponds to the case $\varepsilon_{\mathrm{Reco}}= \varepsilon_{\mathrm{True}}$.
                    The uncertainties of $\varepsilon_{\mathrm{Reco}}$ are on the order of $<3\%$.
			}
  \label{Fig:intercalibration}
\end{minipage}
\end{figure}

As stated in Eq.(\ref{Eq:transparency}), the rate depends also on the hardware state of the instrument quantified by the detection efficiency $\epsilon$ which changes due to the degradation or maintenance activities of various components.
Variations in the reflectivity of optical elements and the responses of camera photodetectors have to be accounted for in dedicated calibration procedures~\cite{CettinaICRC2017}.
Other influence of the hardware is the plan to maintain trigger rates constant by adapting trigger thresholds to the measured individual pixel rates.
This is foreseen for the LSTs, but it is not discussed in this work.

We examined hardware effects using the MC simulations for two telescopes with varying optical efficiencies.
In Fig.~\ref{Fig:Rate_Eff}, the stereo rate exhibits a linear dependence on the product of telescope efficiencies $\varepsilon_{1} \cdot \varepsilon_{2}$ for two configurations $\Omega$ ($5^{\circ}$ and $72^{\circ}$) with fixed $d_{\mathrm{SP}}=150$~m.
This behaviour may be coincidental since it follows from the energy spectrum of cosmic rays that $R \propto E^{-1.7}_{\mathrm{th}}$, where the energy threshold is approximately inversely proportional to the telescope efficiency, i.e~ $E_{\mathrm{th}} \propto \varepsilon^{-1}$.
It further follows that $R \propto \left( \varepsilon_{1} \cdot \varepsilon_{2} \right)^{0.85} = \varepsilon^{1.7}$ for $\varepsilon_{1} = \varepsilon_{2} \equiv \varepsilon$.
This proportionality may differ for instances when $\varepsilon_{1} \neq \varepsilon_{2}$ and it is only the final superposition of all possibilities that behaves linearly.
Considering this linear dependence as an appropriate approximation, for the moment we will assume it as an effective description of the stereo trigger rate.

%***********************************************************************************
% INTER-CALIBRATION
%***********************************************************************************

\vspace{-0.2cm}
\section{Inter-calibration of telescope responses}
\label{Sec:Intercalibration}
\vspace{-0.2cm}

In the following, we will describe the inter-calibration principle for responses of telescopes of the same type using the effective rate correction $F\left( D, \theta, \beta, \vec{B}, \varepsilon \right)$ in Eq.(\ref{Eq:transparency}).

Only stereo rates retrieved for pairs of telescopes are assumed in the calibration.
A cut on the maximum distance of 200~m between telescopes is applied to select such pairwise stereo rates for calibration.
Using the estimates from Eq.(\ref{Eq:rate_fit}) with Table~\ref{Tab:fit} in Eq.(\ref{Eq:transparency}), the rates are corrected for geometrical dependencies  ($D, \theta, \beta$) with the systematic uncertainties, including the effect of the magnetic field, assumed to be within $2\%$ (Sec.~\ref{Sec:Bfield}).
This way, pairwise transparency coefficients $\tau_{ij}$ are obtained for the hardware conditions $\varepsilon_{i} = \varepsilon_{j} = 1$, where $i$ and $j$ label the two telescopes in coincidence.
On the assumption that the atmospheric conditions over the array are uniform, the transparency observed by all telescope pairs is expected to be the same (within the systematic uncertainties), i.e. $\tau_{ij} = \tau_{kl} = T$, where $T$ is the true atmospheric transparency.

Possible degradations of the instrument performance are not included in the definition of $\tau_{ij}$.
As shown in Sec.~\ref{Sec:Hardware}, trigger rates modified by hardware changes can be expressed as $R_{ij} (\varepsilon_{i}, \varepsilon_{j}) \approx \varepsilon_{i} \cdot \varepsilon_{j} \cdot R_{ij} (\varepsilon_{i} = \varepsilon_{j} = 1)$, implying $\tau_{ij} (\varepsilon_{i}, \varepsilon_{j}) \approx \varepsilon_{i} \cdot \varepsilon_{j} \cdot T$.
Variations of telescope efficiencies from the nominal values are then quantified by the asymmetry in coefficients $\tau_{ij}$:
\begin{equation}
    \label{Eq:asymmetry}
        a_{ij/kl} = \frac{\tau_{ij}-\tau_{kl}}{\tau_{ij}+\tau_{kl}}.
\end{equation}

Efficiencies $\varepsilon_i$ are treated as free parameters allowing the relative inter-calibration of telescope responses by means of minimisation of the sum of squared residuals
\begin{equation}
    \label{Eq:chi2}
        \chi^{2} = \sum_{\mathrm{pairs}}{ \left( a_{ij/kl} - \frac{\varepsilon_{i} \cdot \varepsilon_{j} - \varepsilon_{k} \cdot \varepsilon_{l}}{\varepsilon_{i} \cdot \varepsilon_{j} + \varepsilon_{k} \cdot \varepsilon_{l}} \right)^{2} \cdot \sigma^{-2}_{ij/kl}} ,
\end{equation}
where $\sigma^{2}_{ij/kl}$ are the variances of asymmetries $a_{ij/kl}$ and the sum runs over all selected pairs of telescopes present during the data acquisition.

In this work, the described procedure is applied only to the telescopes of the same sub-system.
In the simulated CTA-N layout 3AL4M15-5F, each of the 15 MSTs and 4 LSTs has at least two neighbours within 200~m, providing in total 24 and 6 pairs, respectively.
As the inter-calibration is performed in a relative manner, it requires to fix the value of efficiency of one telescope.
Neither the choice of the reference telescope nor the exact value of the fixed efficiency is relevant for the inter-calibration.

The outlined method was applied to the full CTA-N array~\cite{HassanICRC2017}.
The atmospheric transparency~$T$ was constant in the simulations.
All telescopes got randomly assigned optical efficiencies from the normal distribution $\mathcal{N} (0.7, 0.1)$.
Sets of telescope efficiencies were reconstructed from the minimisation in Eq.(\ref{Eq:chi2}) per each sub-system and then compared to the initial MC values (Fig.~\ref{Fig:intercalibration}).
Both true ($\varepsilon_{\mathrm{True}}$) and recovered ($\varepsilon_{\mathrm{Reco}}$) sets of efficiencies were normalised by their respective mean values.
Their RMS is $\sim3\%$ for the inter-calibration of both the MST and LST sub-systems.

%***********************************************************************************
% ATMOSPHERIC CALIBRATION
%***********************************************************************************

\vspace{-0.3cm}
\section{Atmospheric calibration}
\label{Sec:Atmosphere}
\vspace{-0.2cm}

The calibration of the atmospheric transparency to Cherenkov light is achieved using the relative telescope efficiencies $\varepsilon_{\mathrm{Reco}}$ (see Sec.~\ref{Sec:Intercalibration}). % so that the rate is given as $R = R(\mathrm{AOD})$.
As the absolute value of the normalisation of efficiencies is not specified in this way, it has to be fixed by another calibration procedure~\cite{CettinaICRC2017}.
For the purpose of online monitoring, the normalisation can be chosen from the previous observation run\footnote{Observation run refers to the observation time unit applied in current IACT systems ($\approx 20 - 30$ min).}, assuming these efficiencies do not change significantly within the same night ($<5\%$).

Using the data set from the previous section, we re-scaled each $\varepsilon_{i}$ obtained in the calibration by a factor $\overline{\varepsilon_{\mathrm{True}}} / \overline{\varepsilon_{\mathrm{Reco}}}$.
Inverse values of the re-scaled efficiencies were used to correct transparency coefficients $\tau_{ij}$. 
The mean value of corrected $\tau_{ij}$ for all selected pairs of telescopes determines the estimate of the atmospheric transparency in the observation run $T (\mathrm{AOD}) = 1.02 \pm 0.04$, consistent with the fact that all simulations in this study assumed the same atmospheric conditions.

%***********************************************************************************
% CONCLUSIONS
%***********************************************************************************

\vspace{-0.2cm}
\section{Conclusions}
\label{Sec:Conclusions}
\vspace{-0.2cm}

The impact of atmospheric conditions on the trigger rates and reconstructed effective areas of IACTs has been previously addressed by different methods \cite{Dorner, CAT, Nolan}.
In this work, we have presented another approach using the Cherenkov Transparency Coefficient, successfully implemented in the H.E.S.S. experiment \cite{Hahn}, as an atmospheric sensitive quantity in CTA.

Necessitated by the more complex CTA layout, an effective correction of the trigger rate for the geometrical and hardware effects has been found in order to maintain the CTC independence from these issues.
Neglecting the effects of the magnetic field, the systematic uncertainty of the trigger rate has been found to be $\sim2\%$ (for $d_{\mathrm{SP}}<200$~m).
For the simulated configuration, the recovered CTC is consistent with the expectation.
In addition, the comparison of transparency coefficients obtained per telescope pairs has been shown to be a viable inter-calibration procedure for relative telescope responses with the resolution of reconstructed efficiencies being~$\sim3\%$ for the CTA-N.
All results in this study have been obtained assuming fixed atmospheric conditions.

Since the CTC is calculated using the output of scientific observations, it provides a crosscheck for other calibration methods~\cite{CettinaICRC2017} without the need for additional devices or interference with the regular data taking.
Currently under investigation are the robustness of the CTC under different aerosol concentrations and air density profiles and its application for other telescope types anticipated in CTA (especially for CTA-S) as well as the cross-calibration of different sub-systems.
Future study is also foreseen to investigate the feasibility of the method under varying trigger thresholds of the LSTs.

%***********************************************************************************
% ACKNOWLEDGEMENTS
%***********************************************************************************

\vspace{-0.2cm}
\acknowledgments
\vspace{-0.2cm}
This work was conducted in the context of the CTA Central Calibration Facilities Work Package. 
We gratefully acknowledge financial support from the agencies and organizations listed here: http://www.cta-observatory.org/consortium\_acknowledgments

%***********************************************************************************
% BIBLIOGRAPHY
%***********************************************************************************

%***********************************************************************************
% END DOCUMENT
%***********************************************************************************

\end{document}